\documentclass{article}
\usepackage{amsfonts}
\usepackage{amsmath}
\usepackage{graphicx}
\usepackage{subfigure}
\usepackage{bm}
\usepackage{graphicx}
\usepackage{lineno,hyperref}
\usepackage{amsfonts}
\usepackage{amsmath}
\usepackage{graphicx}
\usepackage{setspace}
\usepackage{float}
\usepackage{caption}
\usepackage[export]{adjustbox}
\usepackage{tabularx}
\usepackage{subfigure}
\usepackage{bm}
\usepackage{listings}
\usepackage{url}
\usepackage{natbib}

{\lstset{language=R,%
 basicstyle=\small\ttfamily,
 showstringspaces=false
 numberstyle=\tiny,
 stepnumber=2,
 frame=lines
 }
 \newcommand{\pkg}[1]{{\normalfont\fontseries{b}\selectfont #1}}
\let\proglang=\textsf
\let\code=\texttt

\begin{document}

\title{\pkg{MaxSkew} and \pkg{MultiSkew}: Two \proglang{R} Packages for
Detecting, Measuring and Removing \\Multivariate Skewness
}


\author{Cinzia Franceschini \and Nicola Loperfido
}


\author{Cinzia Franceschini\\
              Universit\`{a} degli Studi della Tuscia\\Dipartimento di Scienze Agrarie e Forestali (DAFNE) \\
              Via San Camillo de Lellis snc, 01100 Viterbo (VT), ITALY\\
              email: cinziafranceschini@msn.com           \\
Nicola Loperfido \\
              Universit\`{a} degli Studi di Urbino ``Carlo Bo"\\
Dipartimento di Economia, Societ\`{a} e Politica (DESP)\\
Via Saffi 42, 61029 Urbino (PU), ITALY\\
email: nicola.loperfido@uniurb.it
}


\maketitle

\begin{abstract}
Skewness plays a relevant role in several multivariate statistical techniques.
Sometimes it is used to recover data features, as in cluster
analysis. In other circumstances, skewness impairs the performances of
statistical methods, as in the Hotelling's one-sample test. In both
cases, there is the need to check the symmetry of the underlying
distribution, either by visual inspection or by formal testing. The \proglang{R}
packages \pkg{MaxSkew} and \pkg{MultiSkew} address these issues by
measuring, testing and removing skewness from multivariate data. Skewness is
assessed by the third multivariate cumulant and its functions. The hypothesis of symmetry is tested either
nonparametrically, with the bootstrap, or parametrically, under the
normality assumption. Skewness is removed or at least alleviated by projecting
the data onto appropriate linear subspaces. Usages of \pkg{MaxSkew} and \pkg{MultiSkew} are
illustrated with the Iris dataset.\\
\textbf{Keywords:} Asymmetry; Bootstrap; Projection Pursuit; Symmetrization; Third Cumulant.
\end{abstract}

\section{Introduction}
\label{intro}

Skewness of a random variable $X$ satisfying $E\left( \left\vert
X\right\vert ^{3}\right) <+\infty $ is often measured by its third
standardized cumulant
\begin{equation}
\gamma _{1}\left( X\right) =\frac{E\left[ \left( X-\mu\right) ^{3}\right] }{\sigma ^{3}},
\end{equation}
where $\mu $ and $\sigma
$ are the mean and the standard deviation of $X$. The squared third
standardized cumulant $\beta _{1}\left( X\right) =\gamma _{1}^{2}\left(
X\right) $, known as Pearson's skewness, is also used. The numerator of $%
\gamma _{1}\left( X\right) $, that is
\begin{equation}
\kappa_{3}\left( X\right)=E\left[ \left( X-\mu \right) ^{3}%
\right] ,
\end{equation}
is the third cumulant (i.e. the third central moment) of $X$.
Similarly, the third moment (cumulant) of a random vector is a matrix
containing all moments (cumulants) of order three which can be obtained
from the random vector itself.\\
Statistical applications of the third moment
include, but are not limited to: factor analysis (\citealp{BonhommeRobin2009}; \citealp{Mooijaart1985}), density approximation
(\citealp{ ChristiansenLoperfido2014}; \citealp{ Loperfido2019}; \citealp{VanHulle2005}), independent
component analysis \citep{PaajarviLeblanc2004}, financial
econometrics (\citealp{DeLucaLoperfido2015}; \citealp{ElyasianiMansur2017}), cluster analysis (\citealp{KabanGirolami2000}; \citealp{Loperfido2013}; \citealp{Loperfido2015a};
\citealp{Loperfido2019}; \citealp{TarpeyLoperfido2015}), Edgeworth expansions\\ (\citealp{KolloRosen2005},
page 189), portfolio theory
(\citealp{JondeauRockinger2006}), linear models
(\citealp{Mardia1971}; \citealp{YinCook2003}), likelihood inference (\citealp{McCullaghCox1986}), projection pursuit
(\citealp{Loperfido2018}), time series (\citealp{DeLucaLoperfido2015}; \citealp{Fiorentinietal2016}), spatial statistics (\citealp{GentonHeLiu2001};  \citealp{KimMallick2003}; \citealp{Lark2015}). \\
The third cumulant of a $d-$dimensional random vector is a $d^{2}\times d$
matrix with at most $d\left( d+1\right) \left( d+2\right) /6$ distinct
elements. Since their number grows very quickly with the vector's dimension,
it is convenient to summarize the skewness of the random vector itself with
a scalar function of the third standardized cumulant, as for example
Mardia's skewness ({\citealp{Mardia1970}}), partial skewness (\citealp{Davis1980}; \citealp{Isogai1983}; \citealp{Morietal1993}) or directional skewness (\citealp{MalkovichAfifi1973}). These measures have been
mainly used for testing multivariate normality, are invariant with respect
to one-to-one affine transformations and reduce to Pearson's skewness in the
univariate case. {\citet{Loperfido2015b}} reviews their main properties and
investigates their mutual connections.
Skewness might hamper the performance of several multivariate statistical
methods, as for example the Hotelling's one-sample test (\citealp{Mardia1970}; \citealp{Everitt1979}; \citealp{Davis1982}). Symmetry is usually pursued
by means of power transformations, which unfortunately suffer from some
serious drawbacks: the
transformed variables are neither affine invariant nor robust to outliers\\
(\citealp{HubertVeeken2008}; \citealp{LinLin2010}). Moreover,
they might not be easily interpretable nor jointly normal. \cite{Loperfido2014,Loperfido2019} addressed these problems with symmetrizing linear
transformations.\\
The \proglang{R} packages \pkg{MaxSkew} and \pkg{MultiSkew} provide an
unified treatment of multivariate skewness by
detecting, measuring and alleviating skewness from multivariate data. Symmetry is
assessed by either visual inspection or formal testing. Skewness is measured
by the third multivariate cumulant and  its scalar functions. Skewness is removed or at least alleviated
by projecting the data onto appropriate linear subspaces. To the best of our knowledge, no statistical packages
compute their bootstrap estimates, the third cumulant and
linear projections alleviating skewness.
The remainder of the paper is organized as follows. Section 2 reviews
the basic concepts of multivariate skewness within the frameworks of third
moments and projection pursuit. It also describes some skewness-related features of the Iris dataset. Section 3
illustrates the package \pkg{MaxSkew}. Section 4 describes the functions of \pkg{MultiSkew} related to symmetrization, third moments and skewness measures. Section 5 contains  some concluding remarks and hints for improving the packages.
\section{Third moment}
\label{sec:1}
The third multivariate moment, that is the third moment of a random vector,
naturally generalizes to the multivariate case the third moment $E\left(
X^{3}\right) $ of a random variable $X$ whose third absolute moment is finite. It is defined as follows, for a $d-$%
dimensional random vector $\mathbf{x}=\left( X_{1},...,X_{d}\right) ^\top$
satisfying $E\left( \left\vert X_{i}^{3}\right\vert \right) <+\infty $, for $%
i=1,...,d$. The third moment of $\mathbf{x}$ is the $d^{2}\times d$ matrix

\begin{equation}
\mathbf{M}_{3,x}=E\left( \mathbf{x}\otimes \mathbf{x}^\top\otimes \mathbf{x}%
\right),
\end{equation}
where   ``$\otimes $" denotes the Kronecker product (see, for
example, \citealp{Loperfido2015b}). In the following, when referring to the
third moment of a random vector, we shall implicitly assume that all
appropriate moments exist.

The third moment $\mathbf{M}_{3,x}$ of $\mathbf{x}=\left(
X_{1},...,X_{d}\right) ^\top$ contains $d^{3}$ elements of the form $\mu
_{ijh}=E\left( X_{i}X_{j}X_{h}\right) $, where $i,j,h=1$, $...$, $d$.
Many elements are equal to each other, due to the identities
\begin{equation}
\mu
_{ijh}=\mu _{ihj}=\mu _{jih}=\mu _{jhi}=\mu _{hij}=\mu _{hji}.
\end{equation}
First, there are
at most $d$ distinct elements $\mu _{ijh}$
where the three indices are equal
to each other: $\mu _{iii}=E\left( X_{i}^{3}\right) $, for $i=1$, $...$, $d$%
. Second, there are at most $d(d-1)$ distinct elements $\mu _{ijh}
$ where only two indices are equal to each other: $\mu _{iij}=E\left(
X_{i}^{2}X_{j}\right) $, for $i,j=1$, $...$, $d$ and $i\neq j$. Third, there
are at most $d\left( d-1\right) \left( d-2\right) /6$ distinct elements $\mu
_{ijh}$ where the three indices differ from each other: $\mu _{ijh}=E\left(
X_{i}X_{j}X_{h}\right) $, for $i,j=1$, $...$, $d$ and $
i\neq j \neq h $. Hence $\mathbf{M}_{3,x}$ contains at most $d\left(
d+1\right) \left( d+2\right) /6$ distinct elements.

Invariance of $\mu _{ijh}=E\left( X_{i}X_{j}X_{h}\right) $ with respect to
indices permutations implies several symmetries in the structure of $\mathbf{%
M}_{3,x}$. First, $\mathbf{M}_{3,x}$ might be regarded as $d$ matrices $%
\mathbf{B}_{i}=E\left( X_{i}\mathbf{xx}^\top\right) $, ($i=1,...,n$) stacked
on top of each other. Hence $\mu {}_{ijh}$ is the element in the $j$-th row and in the $h$-th
column of the $i$-th block $\mathbf{B}_{i}$ of $\mathbf{M}_{3,x}$.
Similarly, $\mathbf{M}_{3,x}$ might be regarded as $d$ vectorized,
symmetric matrices lined side by side: $\mathbf{M}_{3,x}=\left[ vec\left(
\mathbf{B}_{1}\right) ,...,vec\left( \mathbf{B}_{d}\right) \right] $. Also, left
singular vectors corresponding to positive singular values of the third
multivariate moment are vectorized, symmetric matrices (\citealp{Loperfido2015b}).
Finally, $\mathbf{M}_{3,x}$ might be decomposed into the sum of
at most $d$ Kronecker products of symmetric matrices and vectors (\citealp{Loperfido2015b}).\\
Many useful properties of multivariate moments are related to the linear
transformation $\mathbf{y}=\mathbf{Ax}$, where $\mathbf{A}$ is a $k\times d$
real matrix. The first moment (that is the mean) of $\mathbf{y}$ is
evaluated via matrix multiplication only: $E(\mathbf{y})=\mathbf{A\bm{\mu}}$.
The second moment of $\mathbf{y}$ is evaluated using both the matrix multiplication
and transposition: $\mathbf{M}_{2,y}=\mathbf{AM}_{2,x}\mathbf{A}^\top$ where $\mathbf{M}_{2,x}=E\left( \mathbf{xx}^\top\right) $ denotes the second
moment of $\mathbf{x}$. The
third moment of $\mathbf{y}$ is evaluated using the matrix multiplication,
transposition and the tensor product: $\mathbf{M}_{3,y}=\left( \mathbf{A}%
\otimes \mathbf{A}\right) \mathbf{M}_{3,x}\mathbf{A}^\top$ (\citealp{ChristiansenLoperfido2014}). In particular, the third moment of the linear projection $%
\mathbf{v}^\top\mathbf{x}$, where $\mathbf{v}=\left( v_{1},...,v_{d}\right)
^\top$ is a $d-$dimensional real vector, is $\left( \mathbf{v}^\top\otimes
\mathbf{v}^\top\right) \mathbf{M}_{3,x}\mathbf{v}$ and is
a third-order polynomial in the variables $v_{1}$, ..., $v_{d}$.

The third central moment of $\mathbf{x}$, also known as its third cumulant,
is the third moment of $\mathbf{x}-\bm{\mu}$, where $\bm{\mu}$ is
the mean of $\mathbf{x}$:
\begin{equation}
\mathbf{K}_{3,x}=E\left[ \left( \mathbf{x-\bm{\mu} }\right) \otimes \left(
\mathbf{x-\bm{\mu} }\right) ^\top\otimes \left( \mathbf{x-\bm{\mu} }\right) \right].
\end{equation}
It is related to the third moment via the identity
\begin{equation}
\mathbf{K}_{3,x}=\mathbf{M}_{3,x}-\mathbf{M}_{2,x}\otimes {\bm{\mu} }-%
\bm{\mu}\otimes \mathbf{M}_{2,x}-vec\left( \mathbf{M}_{2,x}\right)
\bm{\mu}^\top+2\bm{\mu}\otimes \bm{\mu}^\top\otimes \bm{\mu}.
\end{equation}%
The third cumulant allows for a better assessment of
skewness by removing the effect of location on third-order moments. It becomes a null matrix under central symmetry,
that is when $\mathbf{x}-\bm{\mu}$ and $\bm{\mu}-\mathbf{x}$ are
identically distributed.

The third standardized moment (or cumulant) of the random vector $\mathbf{x}$
is the third moment of $\mathbf{z}=(Z_{1},...,Z_{d})^\top=\mathbf{\Sigma }%
^{-1/2}\left( \mathbf{x}-\bm{\mu}\right) $, where $\mathbf{\Sigma }%
^{-1/2}$ is the inverse of the positive definite square root $\mathbf{\Sigma
}^{1/2}$\ of $\mathbf{\Sigma }=cov\left( \mathbf{x}\right) $, which is
assumed to be positive definite:

\begin{equation}
\mathbf{\Sigma }^{1/2}=\left( \mathbf{%
\Sigma }^{1/2}\right) ^\top, \mathbf{\Sigma }^{1/2}>0 , \text{and }  \mathbf{\Sigma }%
^{1/2}\mathbf{\Sigma }^{1/2}=\mathbf{\Sigma }.
\end{equation}

It is often denoted by $%
\mathbf{K}_{3,z}$\ and is related to $\mathbf{K}_{3,x}$ via the identity%

\begin{equation}
\mathbf{K}_{3,z}=\left( \mathbf{\Sigma }^{-1/2}\otimes \mathbf{\Sigma }%
^{-1/2}\right) \mathbf{K}_{3,x}\mathbf{\Sigma }^{-1/2}.
\end{equation}
The third standardized cumulant is particularly useful for removing the effects of
location, scale and correlations on third order moments. The mean and the variance of $\mathbf{z}$ are invariant with respect to
orthogonal transformations, but the same does not hold for third
moments: $\mathbf{M}_{3,z}$ and $\mathbf{M}_{3,w}$ will in general differ,
if $\mathbf{w}=\mathbf{Uz}$ and $\mathbf{U}$ is a $d\times d$ orthogonal
matrix.

Projection pursuit is a multivariate statistical technique aimed at finding
interesting low-dimensional data projections. It looks for
the data projections which maximize the projection pursuit index, that is a
measure of interestingness. \cite{Loperfido2018} reviews the merits of skewness (i.e. the third standardized cumulant) as a projection pursuit index. Skewness-based projection pursuit is based on the multivariate skewness measure in \citet{MalkovichAfifi1973}. They defined the directional skewness of a random
vector $\mathbf{x}$ as the maximum value $\beta _{1,d}^{D}\left( \mathbf{x}%
\right) $ attainable by $\beta _{1}\left( \mathbf{c}^\top\mathbf{x}\right) $,
where $\mathbf{c}$ is a nonnull, $d-$dimensional real vector and $\beta
_{1}\left( Y\right) $ is the Pearson's skewness of the random
variable $Y$:%
\begin{equation}
\beta _{1,d}^{D}\left( \mathbf{x}\right) = \max_{\mathbf{c}\in\mathbb{S}^{d-1}} \frac{E^{2}\left[ \left( \mathbf{c}^\top\mathbf{x}-\mathbf{c}%
^\top\bm{\mu }\right) ^{3}\right] }{\left( \mathbf{c}^\top\mathbf{\Sigma c}%
\right) ^{3}},
\end{equation}%
with $\mathbb{S}^{d-1}$ being the set of $d{-}$ dimensional real vectors of unit length.
The name directional skewness reminds that $\beta _{1,d}^{D}\left( \mathbf{x}%
\right) $ is the maximum  skewness attainable by a projection of the random
vector $\mathbf{x}$ onto a direction. It admits a simple representation in terms of the
third standardized cumulant:%
\begin{equation}
\max_{\mathbf{c}\in\mathbb{S}^{d-1}}\left[ \left( \mathbf{c}^\top\otimes \mathbf{c}^\top\right)
\mathbf{K}_{3,z}\mathbf{c}\right] ^{2}=\max_{\mathbf{c}\in\mathbb{S}^{d-1}}\;\underset{f,g,i,j,h,k}{\sum }c_{f}c_{g}c_{i}c_{j}c_{h}c_{k}%
\kappa _{fgi}\kappa _{jhk}.
\end{equation}%
Statistical applications of directional skewness include normality testing\\
(\citealp{MalkovichAfifi1973}), point estimation (\citealp{Loperfido2010}), independent component analysis (\citealp{Loperfido2015b}; \citealp{PaajarviLeblanc2004})
 and cluster analysis (\citealp{KabanGirolami2000}; \citealp{Loperfido2013}; \citealp{Loperfido2015a}; \citealp{Loperfido2018}; \citealp{Loperfido2019}; \citealp{TarpeyLoperfido2015}).

 There is a general consensus that an interesting feature, once found, should
be removed (\citealp{Huber1985}; \citealp{Sun2006}). In skewness-based projection pursuit, this means removing
skewness from the data using appropriate linear transformations. A
random vector whose third cumulant is a null matrix is said to be weakly
symmetric. Weak symmetry might be achieved by linear transformations, when the third cumulant of $\mathbf{x}$ is not
of full rank, and its rows belong to the linear space generated by the right
singular vectors associated with its null singular values. More formally,
let $\mathbf{x}$ be a $d-$dimensional random vector whose third cumulant $%
\mathbf{K}_{3,x}$ has rank $d-k$, with $0<k<d$. Also, let $\mathbf{A}$ be a $%
k\times d$ matrix whose rows span the null space of $\mathbf{K}_{3,x}^\top%
\mathbf{K}_{3,x}$. Then the third cumulant of $\mathbf{Ax}$ is a null
matrix (\citealp{Loperfido2014}). Weak symmetry might be achieved even when this assumption is not
satisfied: any random vector with finite third-order moments and at
least two components admits a projection which is weakly symmetric (\citealp{Loperfido2014}).

The appropriateness of the linear transformation purported to remove or
alleviate symmetry might be assessed with measures of multivariate
skewness, which should be significantly smaller in the transformed data than
in the original ones. \citet{Mardia1970} summarized the multivariate skewness of the
random vector $\mathbf{x}$ with the scalar measure
\begin{equation}
\beta _{1,d}^{M}\left( \mathbf{x}\right) =E\left\{ \left[ \left( \mathbf{x}-%
\bm{\mu }\right) ^\top\mathbf{\Sigma }^{-1}\left( \mathbf{y}-\bm{\mu }%
\right) \right] ^{3}\right\} ,
\end{equation}%
where $\mathbf{x}$ and $\mathbf{y}$ are two $d-$dimensional, independent and
identically distributed random vectors with mean $\bm{\mu }$ and
covariance $\mathbf{\Sigma }$. It might be represented as the squared norm of
the third standardized cumulant:
\begin{equation}
\beta _{1,d}^{M}\left( \mathbf{x}\right) =tr\left( \mathbf{K}_{3,z}^\top%
\mathbf{K}_{3,z}\right) =\underset{i,j,h}{\sum }\kappa _{ijh}^{2}.
\end{equation}%
It is invariant with respect to one-to-one affine transformations:

\begin{equation}
\beta_{1,d}^{M}\left( \mathbf{x}\right) =\beta _{1,d}^{M}\left( \mathbf{Ax+b}%
\right) , \text{where }   \mathbf{b}\in \mathbb{R}^{d}, \mathbf{A}\in \mathbb{R}%
^{d}\times \mathbb{R}^{d}, \det \left( \mathbf{A}\right) \neq 0.
\end{equation}
Mardia's skewness is by far the most popular measure of multivariate skewness. Its
statistical applications include multivariate normality testing (\citealp{Mardia1970}) and assessment of robustness of MANOVA statistics (\citealp{Davis1980}).

Another scalar measure of multivariate skewness is
\begin{equation}
\beta _{1,d}^{P}\left( \mathbf{x}\right) =E\left[ \left( \mathbf{x}-\bm{\mu }\right) ^\top\mathbf{\Sigma }^{-1}\left( \mathbf{x}-\bm{\mu }\right)
\left( \mathbf{x}-\bm{\mu }\right) ^\top\mathbf{\Sigma }^{-1}\left(
\mathbf{y}-\bm{\mu }\right) \left( \mathbf{y}-\bm{\mu }\right) ^\top%
\mathbf{\Sigma }^{-1}\left( \mathbf{y}-\bm{\mu }\right) \right] ,
\end{equation}%
where $\mathbf{x}$ and $\mathbf{y}$ are the same as above. It has been
independently proposed by several authors (\citealp{Davis1980}; \citealp{Isogai1983}; \citealp{Morietal1993}). \cite{Loperfido2015b} named it partial skewness to remind that $\beta
_{1,d}^{P}\left( \mathbf{x}\right) $ does not depend on moments of the form $%
E\left( Z_{i}Z_{j}Z_{h}\right) $ when $i$, $j$, $h$ differ from each other,
as it becomes apparent when representing it as a function of the third
standardized cumulant:
\begin{equation}
\beta _{1,d}^{P}\left( \mathbf{x}\right) =vec^\top\left( \mathbf{I}%
_{d}\right) \mathbf{K}_{3,z}^\top\mathbf{K}_{3,z}vec\left( \mathbf{I}%
_{d}\right) =\underset{i,j,h,k}{\sum }\kappa _{iij}\kappa _{hhk}.
\end{equation}%
Partial skewness is by far less popular than Mardia's skewness. Like the latter
measure, however, it has been applied to multivariate normality testing
(\citealp{Henze1997a}; \citealp{Henze1997b}; \citealp{HenzeKlarMeintanis2003}) and to the assessment of the robustness of MANOVA statistics
(\citealp{Davis1980}).

\citet{Morietal1993} proposed to measure the
skewness of the $d-$dimensional random vector $\mathbf{x}$ with the vector $%
\mathbf{\gamma }_{1,d}(\mathbf{x})=E\left( \mathbf{z}^\top\mathbf{zz}\right) $, where $%
\mathbf{z}$ is the standardized version of $\mathbf{x}$. This vector-valued
measure of skewness might be regarded as weighted average of the
standardized vector $\mathbf{z}$, with more weight placed on the outcomes
furthest away from the sample mean. It is location invariant, admits the
representation $\mathbf{\gamma }_{1,d}(\mathbf{x})=\mathbf{K}_{3,z}^\top vec\left( \mathbf{%
I}_{d}\right) $ and its squared norm is the partial skewness of $\mathbf{x}$%
. It coincides with the third standardized cumulant of a
random variable in the univariate case, and with the null $d-$dimensional
vector when the underlying distribution is centrally symmetric. \cite{Loperfido2015a} applied it to model-based clustering.

\begin{figure}[tbph]
\centering
\includegraphics[scale=0.50]{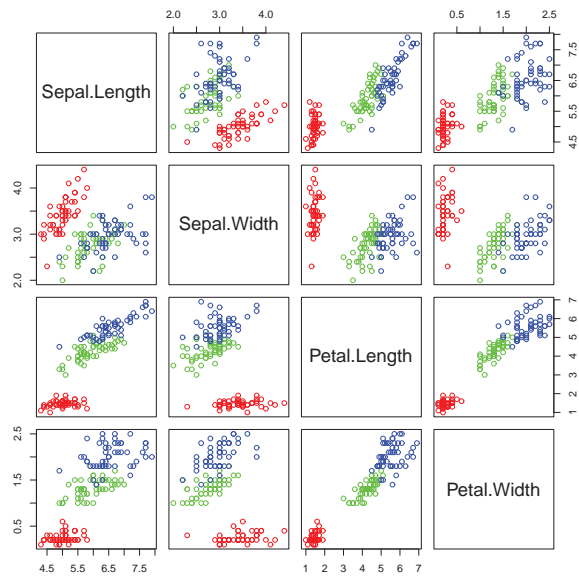}
\caption{Multiple scatterplot of the Iris dataset.}
\label{Fig1}
\end{figure}

We shall illustrate the role of skewness with the
Iris dataset contained in the \textsf{R} package \pkg{datasets}: \code{iris\{datasets\}}. We shall first download the data:
\code{R> data(iris)}.
Then use the help command \code{R>help(iris)}
and the structure command \code{R> str(iris)}
to obtain the following informations about this dataset.
It contains the measurements in centimeters of four variables on 150 iris flowers: sepal length, sepal width, petal length and petal width. There is also a factor variable  (Species) with three levels: setosa, versicolor and virginica. There are 50 rows for each species.
The output of the previous code shows that the dataset is a data frame object containing 150 units and 5 variables:\\
\begin{lstlisting}
'data.frame': 150 obs. of 5 variables:
$ Sepal.Length: num  5.1 4.9 4.7 4.6 5 5.4 4.6 5 4.4 ...
$ Sepal.Width : num  3.5 3 3.2 3.1 3.6 3.9 3.4 3.4 2.9 ...
$ Petal.Length: num  1.4 1.4 1.3 1.5 1.4 1.7 1.4 1.5 1.4 ...
$ Petal.Width : num  0.2 0.2 0.2 0.2 0.2 0.4 0.3 0.2 0.2 ...
$ Species : Factor w/ 3 levels "setosa","versicolor",..: 1 1...
\end{lstlisting}
With the command\\
\code{R> pairs(iris[,1:4],col=c("red","green","blue")[as.numeric(Species)])}
we obtain the multiple scatterplot of the Iris dataset (Figure~\ref{Fig1}).

\begin{figure}[tbph]
\centering%
\subfigure[\protect\url{}\label{fig1bis}]%
{\includegraphics[scale=0.33]{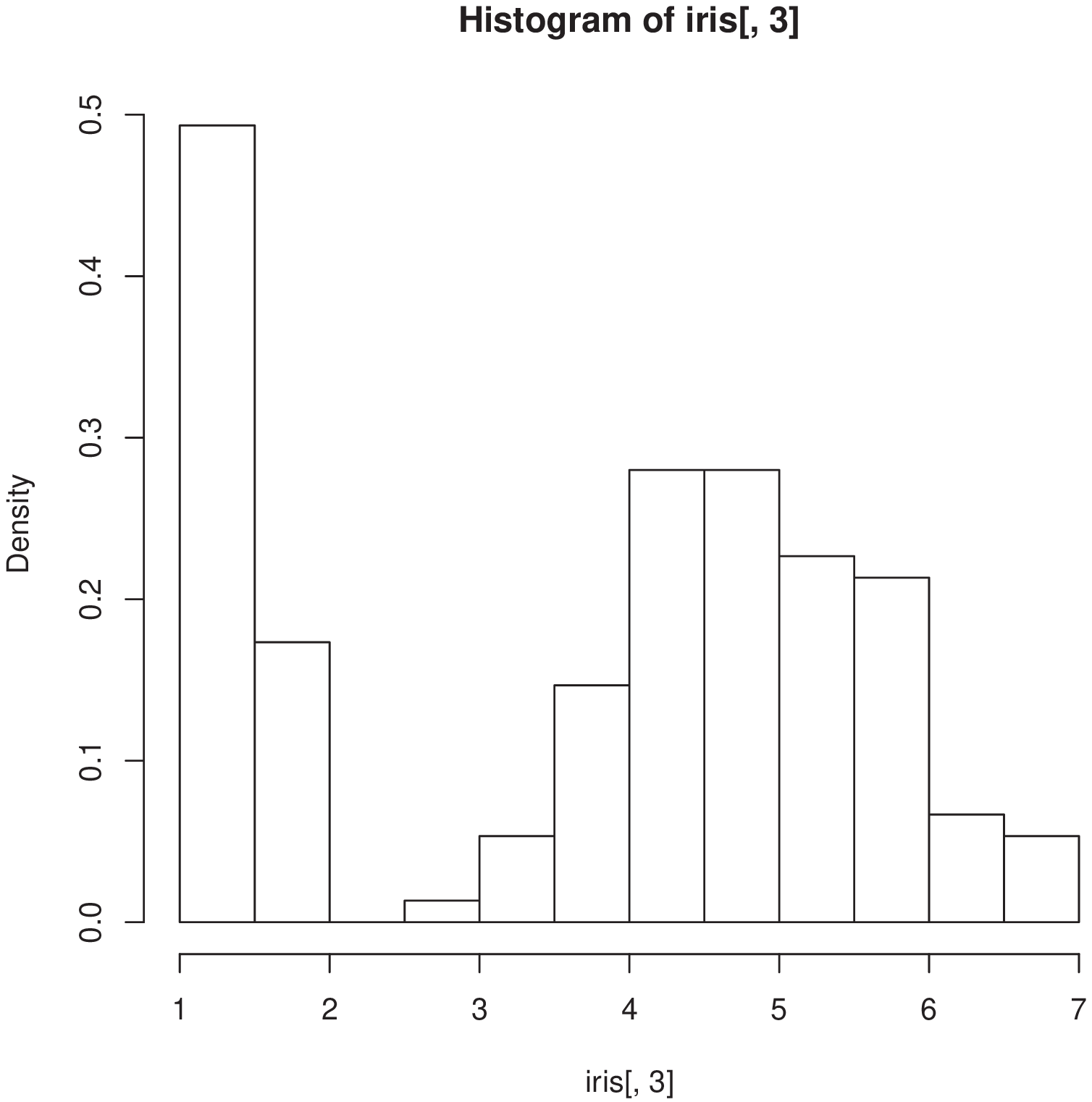}}\qquad\qquad
\subfigure[\protect\url{}\label{fig1ter}]%
{\includegraphics[scale=0.33]{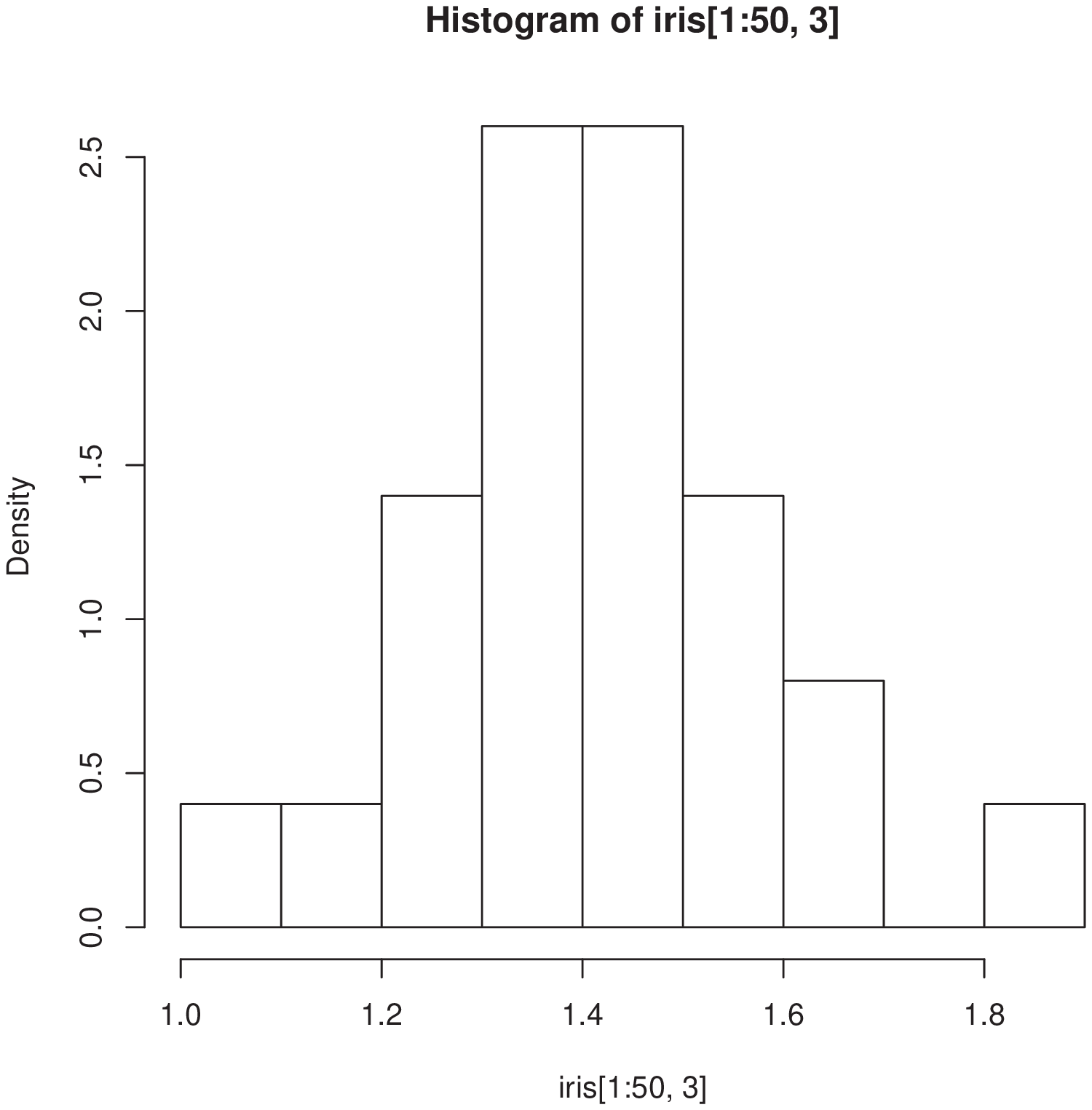}}\qquad\qquad
\caption{(a) Histogram of petal length in the Iris dataset, (b) Histogram of petal length in the ``setosa" group.\label{fig:sottofigure}}

\end{figure}

The groups ``versicolor" and ``virginica" are quite close
to each other, and well separated from ``setosa". The data are markedly
skewed, but skewness reduces within each group, as exemplified by the
histograms of petal length in all flowers (Figure~\ref{fig1bis}) and in those
belonging to the ``setosa" group (Figure~\ref{fig1ter}). Both facts motivated
researchers to model the dataset with mixtures of three normal components
(see, for example \citealp{FrSchnatter2006}, Subsection 6.4.3). However, \citet{KorkmazGoksulukZararsiz2014}
showed that the normality hypothesis should be
rejected at the $0.05$ level in the ``setosa" group, while
nonnormality is undetected by Mardia's skewness.

We shall use \pkg{MaxSkew} and \pkg{MultiSkew} to answer the
following questions about the Iris dataset.
\begin{enumerate}
\item Is skewness really inept at detecting nonnormality of the variable
recorded in the ``setosa" group?

\item Does skewness help in recovering the cluster structure, when information about the group
memberships is removed?

\item Can skewness be removed via linear projections, and are the
projected data meaningful?
\end{enumerate}

The above questions will be addressed within the frameworks of projection
pursuit, normality testing, cluster analysis and data exploration.

\section{MaxSkew}
\label{sec:2}
The package \pkg{MaxSkew} \citep{FranceschiniLoperfido2017a} is written in the \proglang{R}
programming language \citep{rcore}. It is available for download on the Comprehensive \code{R} Archive Network (CRAN) at \\

\url{https://CRAN.R-project.org/package=MaxSkew}.\\

The package \pkg{MaxSkew} uses several \proglang{R} functions. The first one is $\code{eigen}\left \{\code{base}\right \}$  \citep{rcore}. It computes eigenvalues and eigenvectors of real or complex matrices. Its usage, as described by the command \code{help(eigen)} is\\

 \code{R> eigen(x, symmetric, only.values = FALSE, EISPACK = FALSE)}.\\
 \linebreak
The output  of the function are \code{values} (a vector containing the eigenvalues of $x$, sorted in decreasing order according to their modulus) and  \code{vectors} (either a matrix whose columns contain the normalized eigenvectors of $x$, or a null matrix if \code{only.values} is set equal to TRUE). A second  \proglang{R} function used  in \pkg{MaxSkew} is  $\code{polyroot}\left \{\code{base}\right \}$ \citep{rcore} which  finds the zeros of a real or complex polynomial. Its usage is  \code{polyroot(z)}, where $z$ is the vector of polynomial coefficients arranged in increasing order. The \code{Value} in output is a complex vector of length $n - 1$, where $n$ is the position of the largest non-zero element of $z$.  \pkg{MaxSkew} also uses the \proglang{R} function  $\code{svd}\left \{\code{base}\right \}$ \citep{rcore} which computes the singular value decomposition of a rectangular matrix. Its usage is\\

 \code{svd(x, nu = min(n, p), nv = min(n, p), LINPACK = FALSE)}.\\
 \linebreak
The last \proglang{R} function used in \pkg{MaxSkew} is  $\code{kronecker}\left \{\code{base}\right \}$ \citep{rcore} which computes the generalised Kronecker product of two arrays, X and Y.  The \code{Value} in output is   an array  with dimensions \code{dim(X) * dim(Y)}.

 The \pkg{MaxSkew} package finds orthogonal data projections with maximal skewness. The first data projection in the output is the most skewed among all data projections. The second data projection in the output is the most skewed among all data projections orthogonal to the first one, and so on. \cite{Loperfido2019} motivates this method within the framework of model-based clustering. The package implements the algorithm described in \cite{Loperfido2018} and may be downloaded with the command \code{R> install.packages("MaxSkew")}.
The package is attached with the command \code{R > library(MaxSkew)}.\\
The packages \pkg{MaxSkew} and \pkg{MultiSkew} require the dataset to be  a data matrix object, so we transform the  \code{iris} data frame accordingly:\\
\code{R > iris.m<-data.matrix(iris)}. We check that we have a data matrix object with the command:\\
\begin{lstlisting}
R > str(iris.m)
num [1:150, 1:5] {5.1} {4.9} {4.7} {4.6 } {5} {5.4} {...}
- attr(*, "dimnames")=List of 2
..$ : NULL
..$ : chr [1:5] "Sepal.Length" "Sepal.Width"  "Petal.Length"
     "Petal.Width" ...
\end{lstlisting}

The \code{help} command shows the basic informations about the package and the functions it contains: \code{R > help(MaxSkew)}.\\
The package \pkg{MaxSkew} has three functions, two of which are internal.
The usage of the main function is\\
\code{R> MaxSkew(data, iterations, components, plot)},\\
where
\code{data} is a data matrix object, \code{iterations} (the number of required iterations) is a positive integer, \code{components} (the number of orthogonal projections maximizing skewness) is a positive integer smaller than the number of variables, and  \code{plot} is a dichotomous variable: TRUE/FALSE. If \code{plot} is set equal to TRUE (FALSE) the scatterplot of the projections maximizing skewness appears (does not appear) in the output.
The output includes
a matrix of projected data, whose $i$-th row represents the $i$-th unit, while the $j$-th column represents the $j$-th projection. The output also includes
the multiple scatterplot of the projections maximizing skewness. As an example, we call the function\\
\linebreak
\code{R > MaxSkew(iris.m[,1:4], 50, 2, TRUE)}.\\
\linebreak
We have used only the first four columns of the \code{iris.m} data matrix object, because the last column is a label.
As a result, we obtain a matrix with 150 rows and 2 columns containing the projected data and a multiple scatterplot. The structure of the resulting matrix  is\\
\begin{lstlisting}
R> str(MaxSkew(iris.m[,1:4], 50, 2, TRUE))
num [1:150, 1:2] -2.63 -2.62 -2.38 -2.58 -2.57 ...
\end{lstlisting}
For the sake of brevity we only show  the first three rows in the matrix of projected data:\\
\begin{lstlisting}
R > iris.projections<-MaxSkew(iris.m[,1:4], 50,2,TRUE)
R > iris.projections[1:3,]
                [,1]            [,2]
[1,]         -2.631244186      -0.817635353
[2,]         -2.620071890      -1.033692782
[3,]         -2.376652037      -1.311616693
\end{lstlisting}

\begin{figure}[tbph]
\centering%
\subfigure[\protect\url{}\label{Fig2}]%
{\includegraphics[scale=0.33]{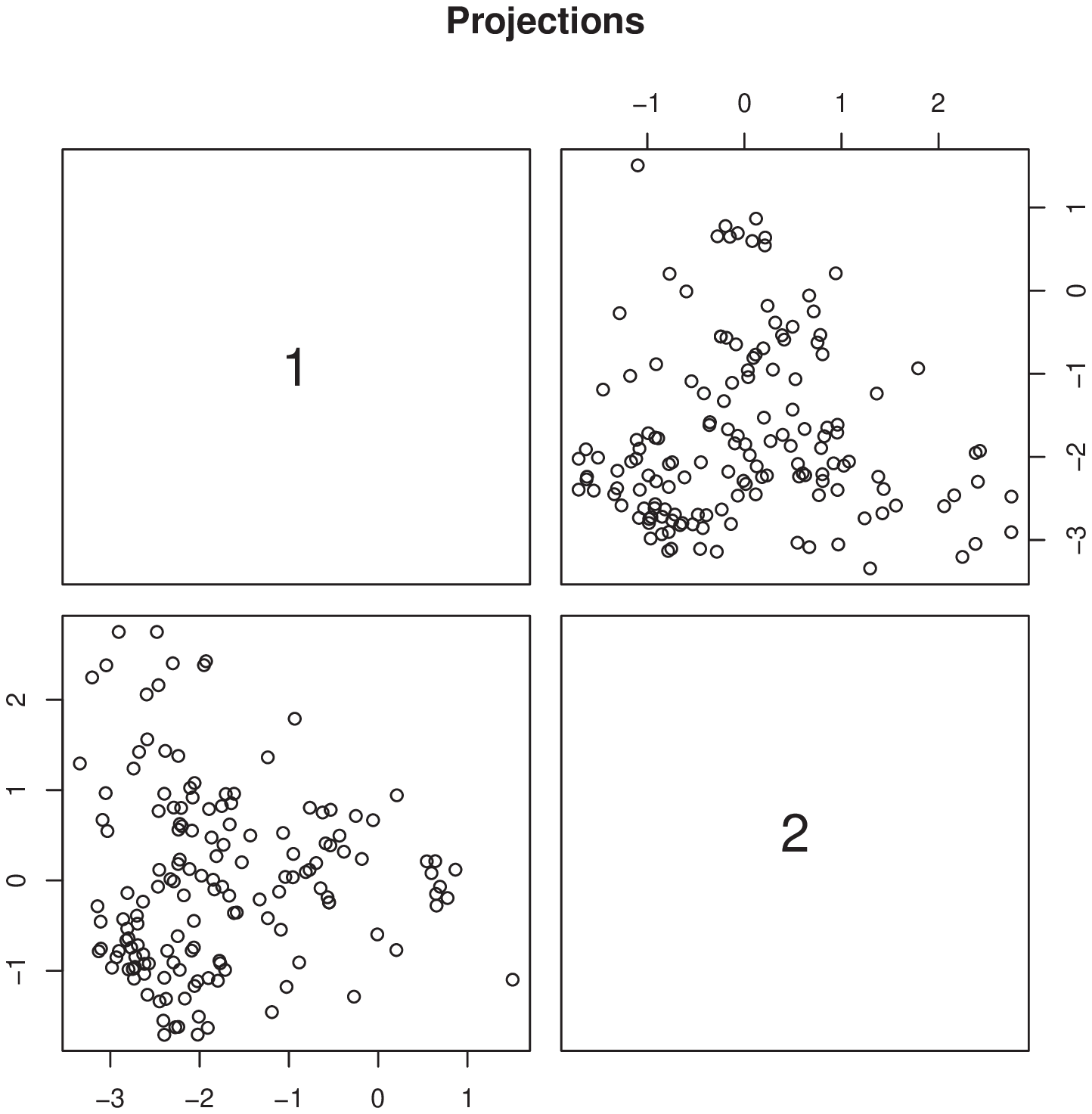}}\qquad\qquad
\subfigure[\protect\url{}\label{Fig3}]%
{\includegraphics[scale=0.33]{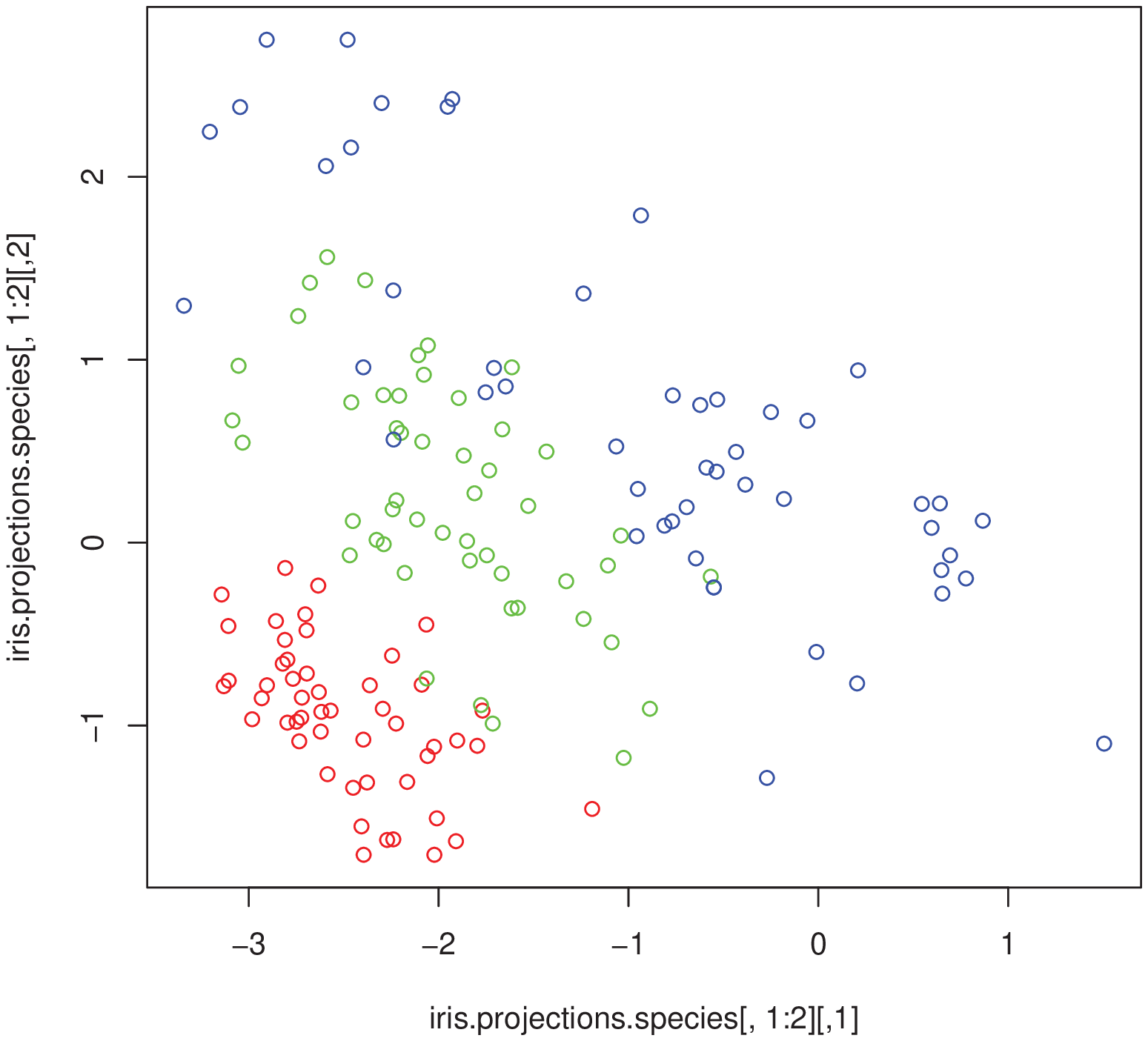}}\qquad\qquad
\caption{(a) Multiple scatterplot of the first two most skewed, mutually orthogonal projections, (b) Scatterplot of the first two most skewed, mutually orthogonal projections of Iris data, with different colors to denote different group memberships.\label{fig:sottofigure2}}
\end{figure}

Figure~\ref{Fig2} shows the scatterplot of the projected data.
The connections between skewness-based projection
pursuit and cluster analysis, as implemented in \pkg{MaxSkew}, have been
investigated by several authors (\citealp{KabanGirolami2000}; \citealp{Loperfido2013}; \citealp{Loperfido2015a}; \citealp{Loperfido2018}; \citealp{Loperfido2019};\\ \citealp{TarpeyLoperfido2015}). For the Iris
dataset, it is well illustrated by the scatterplot of the two most skewed,
mutually orthogonal projections, with different colors to denote the group
memberships (Figure~\ref{Fig3}). The plot is obtained with the following commands:\\
\begin{lstlisting}
R> attach(iris)
R> iris.projections<-MaxSkew(iris.m[,1:4], 50,2,TRUE)
R> iris.projections.species<-cbind(iris.projections,
iris$Species)
R> pairs(iris.projections.species[,1:2],
col=c("red","green","blue")[as.numeric(Species)])
R>detach(iris)
\end{lstlisting}

The scatterplot clearly shows the separation of ``setosa" from
``virginica" and ``versicolor", whose overlapping is much less marked than in
the original variables. The scatterplot is very similar, up to rotation and
scaling, to those obtained from the same data by \cite{FriedmanTukey1974}
and \cite{HuiLindsay2010}.

\begin{figure}[tbph]
\centering
\includegraphics[scale=0.750]{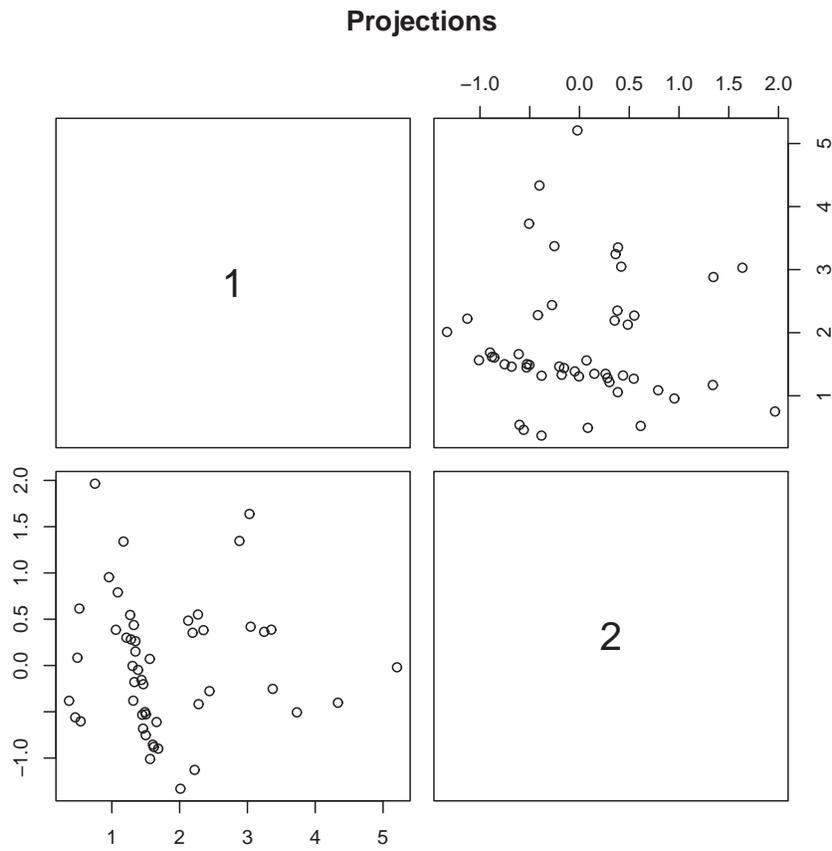}
\caption{Scatterplot of the two most skewed, mutually orthogonal projections computed from data in the ``setosa" group.}
\label{Fig4}
\end{figure}

Mardia's skewness is unable to
detect nonnormality in the ``setosa" group, thus raising the question of
whether any skewness measure is apt at detecting such nonnormality (\citealp{KorkmazGoksulukZararsiz2014}). We shall
address this question using skewness-based projection pursuit as a
visualization tool. Figure~\ref{Fig4} contains the scatterplot of the two most
skewed, mutually orthogonal projections obtained from the four variables
recorded from setosa flowers only. It clearly shows the presence of a dense,
elongated cluster, which is inconsistent with the normality assumption.
Formal testing for excess skewness in the ``setosa" group will be discussed
in the following sections.\\

\section{MultiSkew}
\label{sec:3}
The package \pkg{MultiSkew}  (\citealt{FranceschiniLoperfido2017b}) is written in the \proglang{R}
programming language \citep{rcore} and depends on the recommended
package \pkg{MaxSkew} (\citealt{FranceschiniLoperfido2017a}).
It is available for download on the Comprehensive \proglang{R} Archive Network (CRAN) at {\url{https://CRAN.R-project.org/package=MultiSkew}}.
The \pkg{MultiSkew} package computes the third multivariate cumulant of either the raw, centered or standardized data. It also computes the main measures of multivariate skewness, together with their bootstrap distributions. Finally, it computes the least skewed linear projections of the data. The \pkg{MultiSkew} package contains six different functions. First install it with the command\\
\linebreak
\code{R> install.packages("MultiSkew")}
\linebreak
\\
and then use the command
\code{R > library(MultiSkew)} to attach the package.
Since the package \pkg{MultiSkew} depends on the package \pkg{MaxSkew}, the latter is loaded together to the former.\\
\subsection{MinSkew}
\label{sec:4}
\begin{figure}[tbph]
\centering
\includegraphics[scale=0.750]{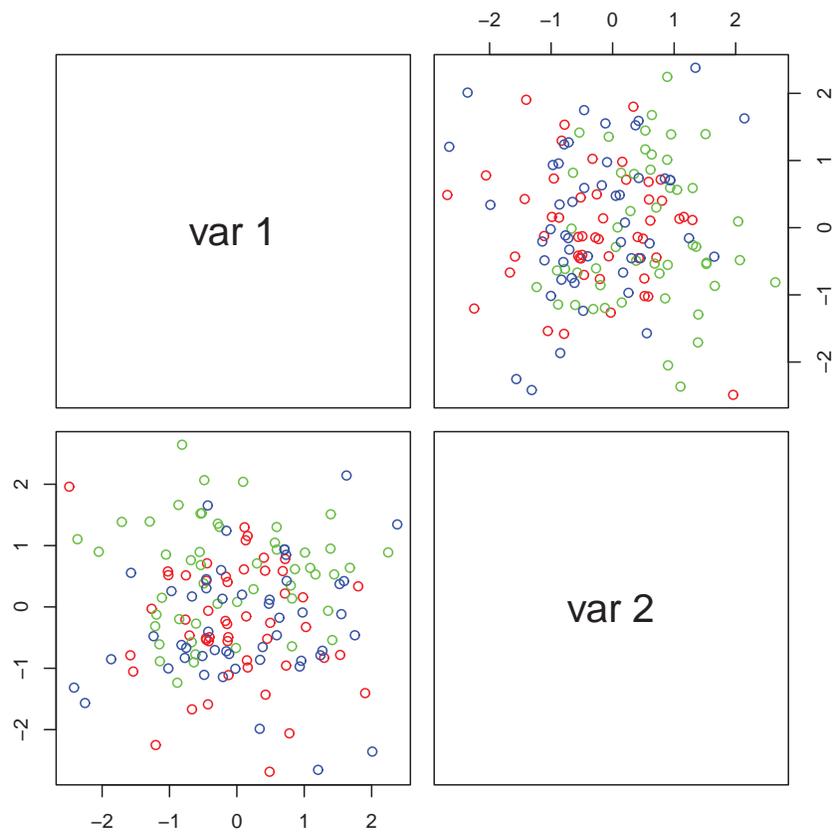}
\caption{Scatterplot of two projections obtained with the function \code{Minskew}.}
\label{Fig5}
\end{figure}

The function \code{R > MinSkew(data, dimension)}
alleviates sample skewness by projecting the data onto appropriate linear subspaces and implements the method in \cite{Loperfido2014}.
It requires two input arguments: \code{data} (a data matrix object), and \code{dimension} (the number of required projections),
which must be an integer between 2 and the number of the variables in the data matrix.
The output has two values: \code{Linear} (the linear function of the variables) and \code{Projections} (the projected data).
\code{Linear} is a matrix with the number of rows and columns equal to the number of variables and number of projections. \code{Projections} is a matrix whose number of rows and columns equal the number of observations and the number of projections.
We call the function using our data matrix object: \code{R > MinSkew(iris.m[,1:4],2)}.\\
We obtain the matrix \code{Linear} as a first output:\\
With the commands
\begin{lstlisting}
R> attach(iris)
R> projections.species<-cbind(Projections,iris$Species)
R> pairs(projections.species[,1:2],col=c("red","green",
"blue")[as.numeric(Species)])
R> detach(iris)
\end{lstlisting}
we obtain the multiple scatterplot of the two projections (Figure~\ref{Fig5}). The points clearly remind of  bivariate normality, and the groups markedly overlap with each other. Hence the projection removes the group structure as well as skewness. This result might be useful for a researcher interested in the  features which are common to the three species.

\begin{figure}[tbph]
\centering%
\subfigure[\protect\url{}\label{Fig6}]%
{\includegraphics[scale=0.33]{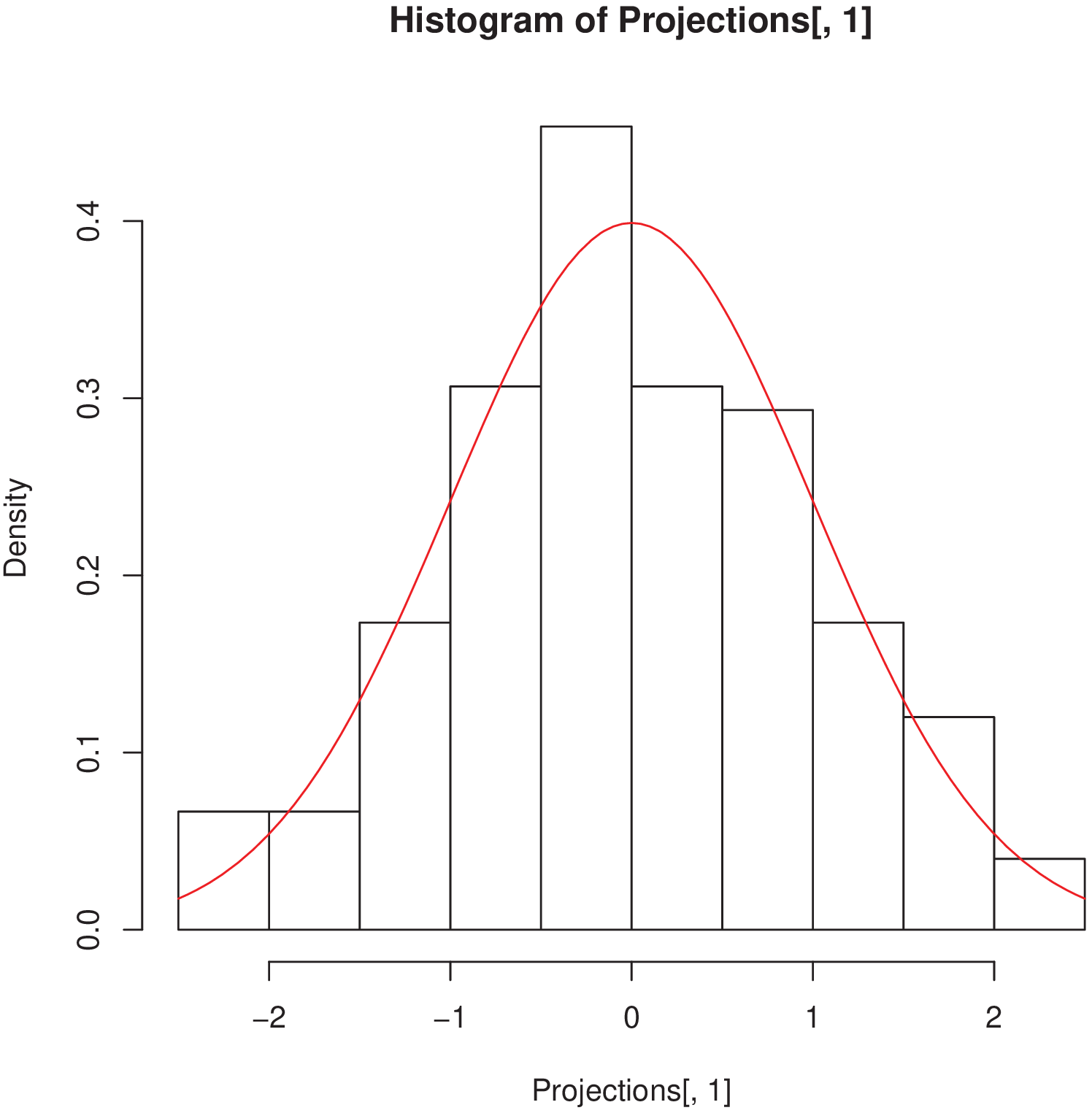}}\qquad\qquad
\subfigure[\protect\url{}\label{Fig7}]%
{\includegraphics[scale=0.33]{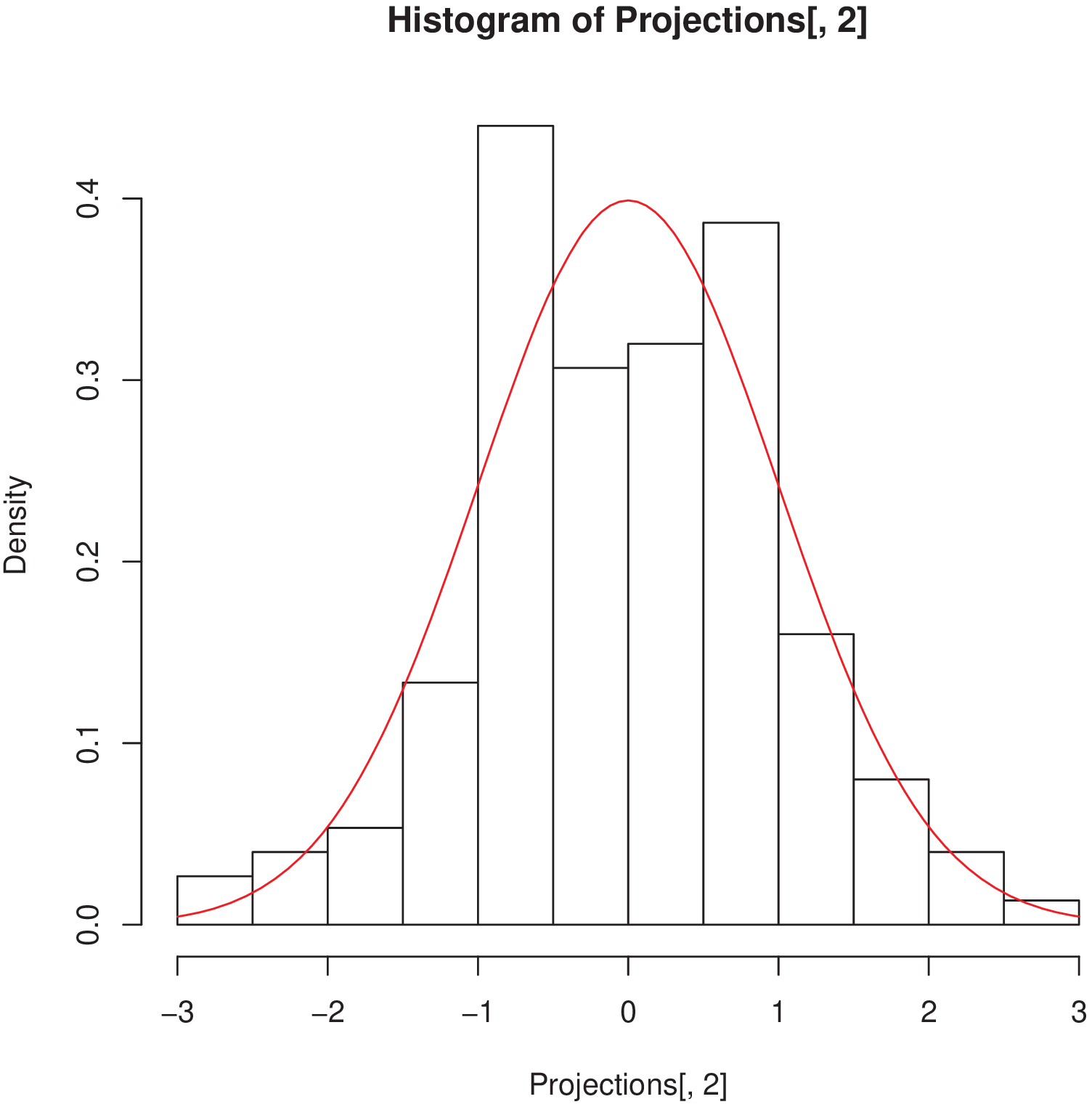}}\qquad\qquad
\caption{(a) Histogram of the first projection of the Iris dataset obtained with \code{MinSkew}, (b) Histogram of the second projection of the Iris dataset obtained with \code{MinSkew}.\label{fig:sottofigure3}}
\end{figure}

The histograms of the \code{MinSkew} projections remind of univariate normality, as it can be seen from  Figure~\ref{Fig6} and Figure~\ref{Fig7}, obtained with the code
\begin{lstlisting}
R> hist(Projections[,1],freq=FALSE)
R> curve(dnorm, col = 2, add = TRUE)
R> hist(Projections[,2],freq=FALSE)
R> curve(dnorm, col = 2, add = TRUE)
\end{lstlisting}

\subsection{Third}
\label{sec:5}

This section describes the function which computes the third multivariate moment of a data matrix. Some general information about the third multivariate moment of both theoretical and empirical distributions are reviewed in \cite{Loperfido2015b}. The name of the function is \code{Third}, and its usage is\\
\linebreak
\code{R > Third(data,type)}.\\
As before, \code{data} is a data matrix object while \code{type} may be:
``raw" (the third raw moment), ``central" (the third central moment)  and ``standardized" (the third standardized moment).
The output of the function, called \code{ThirdMoment}, is a matrix containing all moments of order three which can be obtained from the variables in \code{data}.
We compute the third raw moments of the iris variables with the command \code{R >  Third(iris.m[,1:4], "raw")}.\\
The matrix \code{ThirdMoment} is:\\
\begin{lstlisting}
[1] "Third Moment"

       [,1]      [,2]       [,3]       [,4]
[1,]  211.6333  106.0231   145.8113   47.7868
[2,]  106.0231   55.4270    69.1059   22.3144
[3,]  145.8113   69.1059   109.9328   37.1797
[4,]   47.7868   22.3144    37.1797   12.9610
[5,]  106.0231   55.4270    69.1059   22.3144
[6,]   55.4270   30.3345    33.7011   10.6390
[7,]   69.1059   33.7011    50.7745   17.1259
[8,]   22.3144   10.6390    17.1259    5.9822
[9,]  145.8113   69.1059   109.9328   37.1797
[10,]  69.1059   33.7011    50.7745   17.1259
[11,] 109.9328   50.7745    86.4892   29.6938
[12,]  37.1797   17.1259    29.6938   10.4469
[13,]  47.7868   22.3144    37.1797   12.9610
[14,]  22.3144   10.6390    17.1259    5.9822
[15,]  37.1797   17.1259    29.6938   10.4469
[16,]  12.9610    5.9822    10.4469    3.7570
\end{lstlisting}
\begin{lstlisting}
R> str(ThirdMoment)
num [1:16, 1:4] 211.6 106 145.8 47.8 106 ...
\end{lstlisting}
Similarly, we use ``central" instead of ``raw" with the command\\
\code{R >  Third(iris.m[,1:4], "central")}.\\
\linebreak
The output which appears in console is\\
\begin{lstlisting}
[1] "Third Moment"

        [,1]     [,2]     [,3]     [,4]
[1,]   0.1752   0.0420   0.1432   0.0259
[2,]   0.0420  -0.0373   0.1710   0.0770
[3,]   0.1432   0.1710  -0.1920  -0.1223
[4,]   0.0259   0.0770  -0.1223  -0.0466
[5,]   0.0420  -0.0373   0.1710   0.0770
[6,]  -0.0373   0.0259  -0.1329  -0.0591
[7,]   0.1710  -0.1329   0.5943   0.2583
[8,]   0.0770  -0.0591   0.2583   0.1099
[9,]   0.1432   0.1710  -0.1920  -0.1223
[10,]  0.1710  -0.1329   0.5943   0.2583
[11,] -0.1920   0.5943  -1.4821  -0.6292
[12,] -0.1223   0.2583  -0.6292  -0.2145
[13,]  0.0259   0.0770  -0.1223  -0.0466
[14,]  0.0770  -0.0591   0.2583   0.1099
[15,] -0.1223   0.2583  -0.6292  -0.2145
[16,] -0.0466   0.1099  -0.2145  -0.0447
\end{lstlisting}
and the structure of the object \code{ThirdMoment} is\\
\begin{lstlisting}
R> str(ThirdMoment)
num [1:16, 1:4] 0.1752 0.042 0.1432 0.0259 0.042 ...
\end{lstlisting}
Finally, we set \code{type} equal to \code{standardized}:\\
\code{R > Third(iris.m[,1:4], "standardized")},\\
\linebreak
and obtain the output\\
\begin{lstlisting}
  [1] "Third Moment"

           [,1]     [,2]     [,3]     [,4]
[1,]      0.2988   -0.0484   0.3257    0.0034
[2,]     -0.0484    0.0927} -0.0358   -0.0444
[3,]      0.3257   -0.0358}  0.0788   -0.2221
[4,]      0.0034   -0.0444} -0.2221    0.0598
[5,]     -0.0484    0.0927} -0.0358   -0.0444
[6,]      0.0927   -0.0331} -0.1166   -0.0844
[7,]     -0.0358   -0.1166}  0.2894    0.1572
[8,]     -0.0444   -0.0844}  0.1572    0.2276
[9,]      0.3257   -0.0358}  0.0788   -0.2221
[10,]    -0.0358   -0.1166}  0.2894    0.1572
[11,]     0.0788    0.2894} -0.0995   -0.3317
[12,]    -0.2221    0.1572} -0.3317    0.3009
[13,]     0.0034   -0.0444} -0.2221    0.0598
[14,]    -0.0444   -0.0844}  0.1572    0.2276
[15,]    -0.2221    0.1572} -0.3317    0.3009
[16,]     0.0598    0.2276}  0.3009    0.8259

\end{lstlisting}
We show the structure of the matrix ThirdMoment with \\
\begin{lstlisting}
R> str(ThirdMoment)
num [1:16, 1:4] 0.2988 -0.0484 0.3257 0.0034 -0.0484 ...
\end{lstlisting}
Third moments and cumulants might give some insights into
the data structure. As a first example, use the command\\
\code{R> Third(MaxSkew(iris.m[1:50,1:4],50,2,TRUE),"standardized")}\\
to compute the third standardized cumulant of the two most skewed, mutually
orthogonal projections obtained from the four variables recorded from setosa
flowers only. The resulting matrix is\\
\begin{lstlisting}
[1] "Third Moment"

       [,1]     [,2]
[1,]  1.2345   0.0918
[2,]  0.0918  -0.0746
[3,]  0.0918  -0.0746
[4,] -0.0746   0.5936
\end{lstlisting}
The largest entries in the matrix are the first element of the first row and
the last element in the last row. This pattern is typical of outcomes from
random vectors with skewed, independent components (see, for example,
\citealt{Loperfido2015b}). Hence the two most skewed projections may well be mutually independent.
As a second example, use the commands\\
\code{R> MinSkew(iris.m[,1:4],2)} and \code{R> Third(Projections,"standardized")},\\
where \code{Projections} is a value in output of the function \code{MinSkew},
to compute the third standardized cumulant of the two least skewed
projections obtained from the four variables of the Iris dataset.
The resulting matrix is\\
\begin{lstlisting}
      [,1]     [,2]
[1,] -0.0219  0.0334
[2,]  0.0334 -0.0151
[3,]  0.0334 -0.0151
[4,] -0.0151 -0.0963
\end{lstlisting}
All elements in the matrix are very close to zero, as it might be better
appreciated by comparing them with those in the third standardized cumulant
of the original variables. This pattern is typical of outcomes from
weakly symmetric random vectors (\citealp{Loperfido2014}).

\subsection{Skewness measures}
\label{sec:6}

\begin{figure}[tbph]
\centering
\includegraphics[scale=0.50]{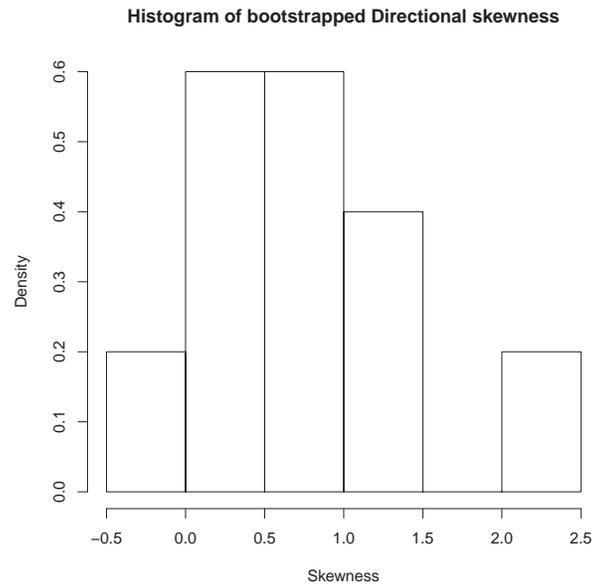}
\caption{Bootstrapped directional skewness of Iris dataset.}
\label{Fig8}
\end{figure}

The package \pkg{MultiSkew} has other four functions. All of them compute skewness measures. The first one is
\code{R > FisherSkew(data)} and  computes Fisher's measure of skewness, that is the third standardized moment of each variable in the dataset.
The usage of the function shows that there is only one input argument: \code{data} (a data matrix object). The output of the function is a dataframe, whose name is \code{tab}, containing Fisher's measure of skewness of each variable of the dataset. To illustrate  the function, we use the four numerical variables in the Iris dataset:\\
\code{R > FisherSkew(iris.m[,1:4])}\\
and obtain the output\\
\begin{lstlisting}
R > tab
                  X1          X2         X3        X4
Variables       1.0000     2.0000     3.0000     4.000
Fisher Skewness 0.3118     0.3158    -0.2721    -0.1019
\end{lstlisting}
in which \code{X1} is the variable Sepal.Length, \code{X2} is the variable Sepal.Width, \code{X3} is the variable Petal.Length, and \code{X4} is the variable Petal.Width.
Another function is \code{PartialSkew: R > PartialSkew(data)}.
It computes the multivariate skewness measure as defined in \cite{Morietal1993}.
The input is still a data matrix, while the values in output are three objects:
\code{Vector, Scalar} and \code{pvalue}.
The first is the skewness measure and it has a number of elements equal to the number of the variables in the dataset used as input.
The second is the squared norm of \code{Vector}. The last is the probability of observing a value of \code{Scalar} greater than the observed one, when the data are normally distributed and the sample size is large enough. We apply this function to our dataset: \code{R > PartialSkew(iris.m[,1:4])} and obtain\\
\begin{lstlisting}
R > Vector
                [,1]
[1,]            0.5301
[2,]            0.4355
[3,]            0.4105
[4,]            0.4131

R > Scalar
                [,1]
[1,]            0.8098

R > pvalue
                [,1]
[1,]            0.0384
\end{lstlisting}
The function \code{R > SkewMardia(data)} computes the multivariate skewness introduced in \cite{Mardia1970}, that is the sum of squared elements in the third standardized cumulant of the data matrix. The output of the function is the squared norm of the third cumulant of the standardized data (\code{MardiaSkewness}) and the probability of observing a value of \code{MardiaSkewness} greater than the observed one, when data are normally distributed and the sample size is large enough (\code{pvalue)}.\\
With the command \code{R > SkewMardia(iris.m[,1:4])} we obtain\\
\begin{lstlisting}
R > MardiaSkewness
[1] 2.69722

R > pvalue
[1] 4.757998e-07
\end{lstlisting}
The function \code{SkewBoot} performs bootstrap inference for multivariate skewness measures.
It computes the bootstrap distribution, its histogram and the corresponding $p$-value of the chosen measure of multivariate skewness using a given number of bootstrap replicates. The function calls the function \code{MaxSkew} contained in \pkg{MaxSkew} package. Here, the number of iterations required by the function \code{MaxSkew} is set equal to 5.
The function's usage is
\linebreak
\code{R > SkewBoot(data, replicates, units, type)}. It requires four inputs:
\code{data} (the usual data matrix object),
\code{replicates} (the number of bootstrap replicates),
\code{units} (the number of rows in the data matrices sampled from the original data matrix object) and
\code{type}. The latter may be ``Directional", ``Partial" or ``Mardia" (three different measures of multivariate skewness). If \code{type} is set equal to ``Directional" or ``Mardia", \code{units} is an integer greater than the number of variables. If \code{type} is set equal to ``Partial", \code{units} is an integer greater than the number of variables augmented by one.
The values in output are three:
\code{histogram} (a plot of the above mentioned bootstrap distribution),
\code{Pvalue} (the $p$-value of the chosen skewness measure) and
\code{Vector} (the vector containing the bootstrap replicates of the chosen skewness measure).
For the reproducibility of the result, before calling the function \code{SkewBoot}, we type
\code{R> set.seed(101)}
and after\\
\code{R > SkewBoot(iris.m[,1:4],10,11,"Directional")}.\\
We obtain the output
\begin{lstlisting}
[1] "Vector"
[1]  2.0898  1.4443  1.0730  0.7690  0.6914  0.3617  0.2375  0.0241
[9] -0.1033  0.6092

[1]  "Pvalue"
[1] 0.7272727
\end{lstlisting}
and also the histogram of bootstrapped directional skewness (Figure~\ref{Fig8}).

We call the function \code{SkewBoot}, first setting \code{type} equal to \code{Mardia} and then equal to \code{Partial}:\\
\begin{lstlisting}
R> set.seed(101)
R > SkewBoot(iris.m[,1:4],10,11,"Mardia").
\end{lstlisting}
We obtain the output  \\
\begin{lstlisting}
[1] "Vector"
[1]  1.4768  1.1260  0.8008  0.6164  0.4554  0.1550  0.0856
    -0.1394 -0.1857  0.4018

[1]  "Pvalue"
[1] 0.6363636

R> set.seed(101)
R > SkewBoot(iris.m[,1:4],10,11,"Partial"),

[1] "Vector"
[1] 1.5435  1.0110  0.6338  0.2858 -0.0053 -0.3235 -0.6701
-1.1134 -1.7075 -0.9563

[1]  "Pvalue"
[1] 0.3636364
\end{lstlisting}

\begin{figure}[tbph]
\centering%
\subfigure[\protect\url{}\label{Fig9}]%
{\includegraphics[scale=0.33]{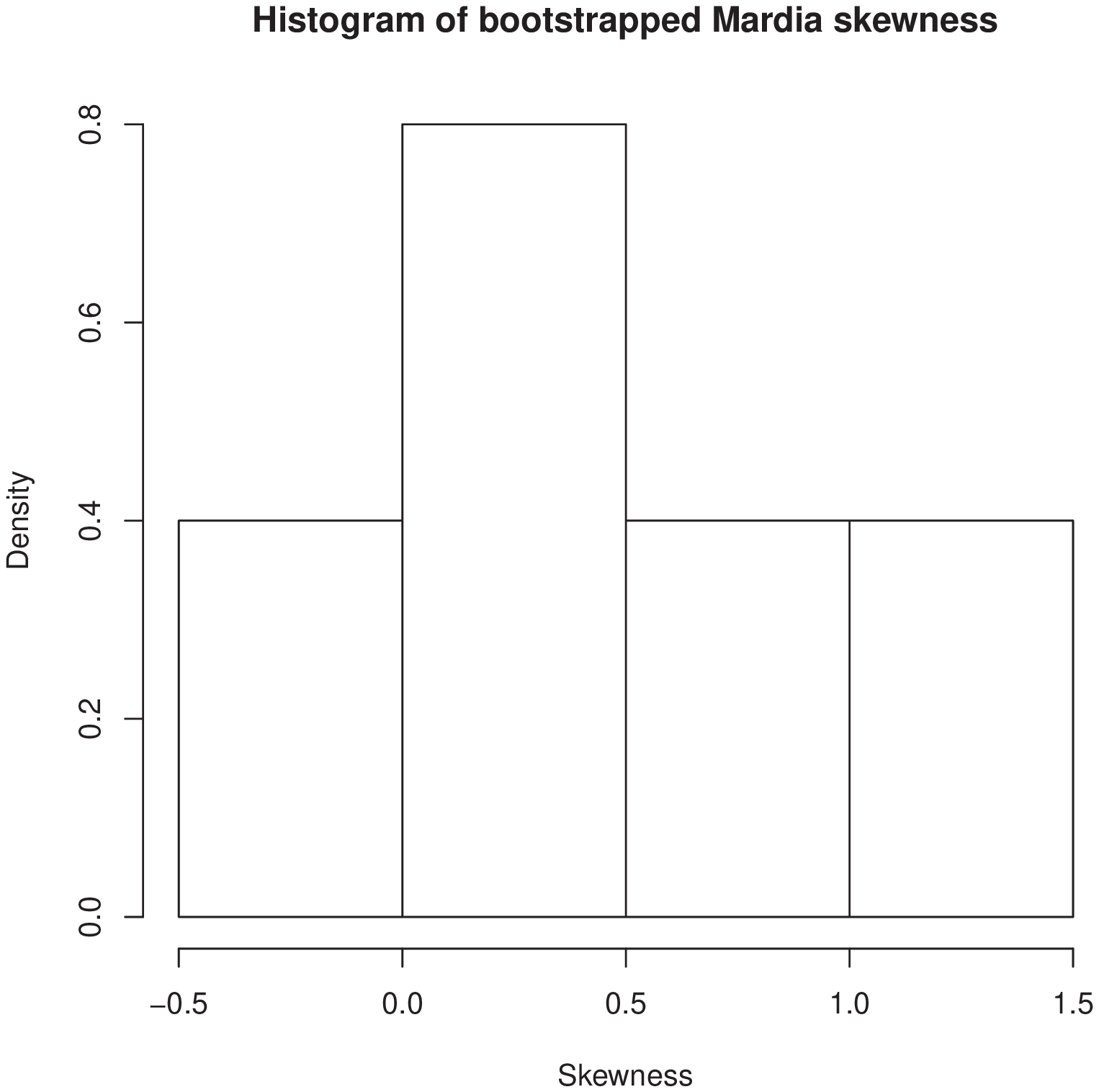}}\qquad\qquad
\subfigure[\protect\url{}\label{Fig10}]%
{\includegraphics[scale=0.33]{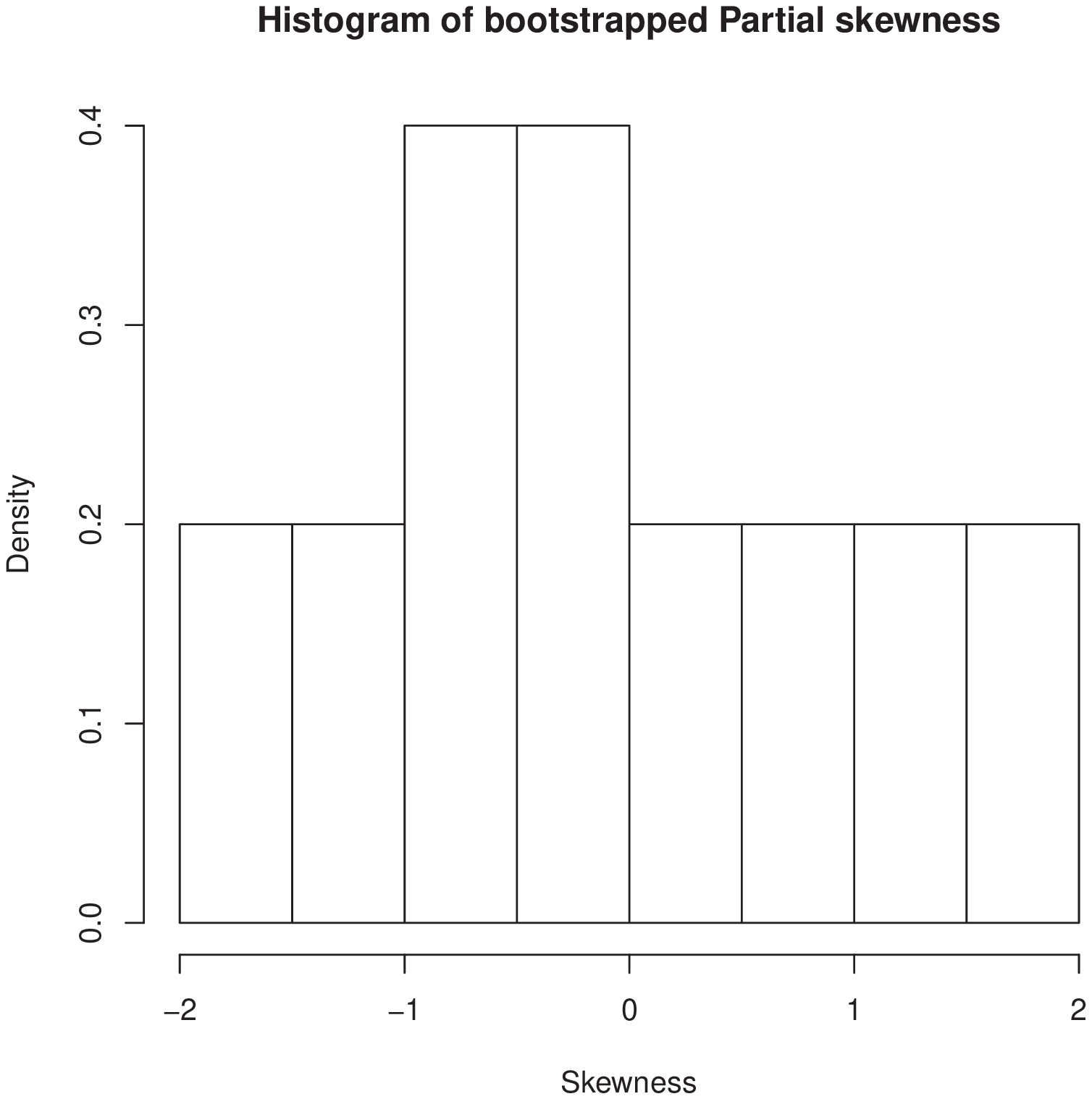}}\qquad\qquad
\caption{(a) Bootstrapped Mardia's skewness of Iris dataset, (b) Bootstrapped Partial skewness of Iris dataset.\label{fig:sottofigure4}}
\end{figure}

Figure~\ref{Fig9} and Figure~\ref{Fig10} contain the histograms of bootstrapped Mardia's skewness and bootstrapped partial skewness, respectively.\\

We shall now compute the Fisher's skewnesses of
the four variables in the ``setosa'' group with\\
\code{R > FisherSkew(iris.m[1:50,1:4])}.
 \linebreak
The output is the dataframe\\
\begin{lstlisting}
R> tab

                    X1          X2        X3          X4
Variables        1.0000      2.0000     3.0000     4.0000
Fisher Skewness  0.1165      0.0399     0.1032     1.2159

\end{lstlisting}
\citet{KorkmazGoksulukZararsiz2014} showed that nonnormality of variables in the
``setosa" group went undetected by Mardia's skewness calculated on all four
of them (the corresponding $p-$value is 0.1772). Here, we shall compute
Mardia's skewness of the two most skewed variables (sepal length and petal
width), with the commands\\
\linebreak
\code{R>iris.m.mardia<-cbind(iris.m[1:50,1],iris.m[1:50,4])}\\
\code{R>SkewMardia(iris.m.mardia)}.\\
\linebreak
We obtain the output\\
\begin{lstlisting}
R> MardiaSkewness
[1] 1.641217

R> pvalue
[1] 0.008401288
\end{lstlisting}
The $p-$value is highly significant, clearly suggesting the presence of
skewness and hence of nonnormality. We conjecture that the normality test
based on Mardia's skewness is less powerful when skewness is present only in
a few variables, either original or projected, while the remaining variables
might be regarded as background noise. We hope to either prove or disprove
this conjecture by both theoretical results and numerical experiments in
future works.
\section{Conclusions}
\label{S:8}

\pkg{MaxSkew} and \pkg{MultiSkew} are two \proglang{R} packages aimed at
detecting, measuring and removing multivariate skewness. They also compute the three
main skewness measures. The function \code{SkewBoot} computes the
bootstrap \textit{p-}value corresponding to the chosen skewness measure.
Skewness removal might be achieved with the function \code{
MinSkew}. The function \code{Third}, which computes the third moment,
plays a role whenever the researcher compares the third sample moment with
the expected third moment under a given model, in order to get a better
insight into the model's fit.

The major drawback of both \pkg{MaxSkew} and \pkg{Multiskew} is that
they address skewness by means of third-order moments only. In the first
place, they may not exist even if the distribution is skewed, as it happens
for the skew-Cauchy distribution. In the second
place, the third moment of a random vector may be a null matrix also when
the random vector itself is asymmetric. In the third place, third-order
moments are not robust to outliers. We are currently investigating these
problems.


\bibliography{All}{}
\bibliographystyle{plainnat}

%
%

\end{document}